# Electron Delocalization versus Emission Coherence of Quantum Dot Superlattices

*Lanfang Hou[1], Zijian He[2], Kexin Wang[3], Kai Wang[2], Shun Wang[1], Butian Zhang[1,*]*

[1]National Gravitation Laboratory, MOE Key Laboratory of Fundamental Physical Quantities Measurement, and School of Physics, Huazhong University of Science and Technology, Wuhan 430074, People's Republic of China

[2]Wuhan National Laboratory for Optoelectronics and School of Physics, Huazhong University of Science and Technology, Wuhan 430074, China

[3]Yunnan Key Laboratory of III-V Infrared Detectors, Kunming Institute of Physics, Kunming 650223, China

ABSTRACT. Cooperative emission is a collective quantum optical process that requires macroscopic phase coherence among coupled emitters. Recent observations of cooperative emission in QD superlattices have renewed interest in how such coherence emerges in nanostructured solids. Meanwhile, theoretical studies have long discussed the relationship between electronic delocalization and coherence, particularly whether delocalized states necessarily give rise to cooperative emission. This study addresses this question through power-dependent steady-state PL and time-resolved PL decay measurements. The findings indicate that, although the quantum resonance peak exhibits delocalized excitonic characteristics, it shows no signatures of cooperative radiation. In particular, neither superlinear intensity scaling nor power-dependent emission delay was observed, indicating the absence of cooperative-radiation signatures. This can be understood from two disorder-related aspects. Temperature-dependent spectroscopy reveals pronounced inhomogeneous broadening and low-temperature dark-exciton participation, pointing to intra-domain static disorder and exciton-state mixing. These effects collectively hinder the establishment of macroscopic coherence. The temperature dependence of the quantum resonance peak decay lifetime is consistent with two-dimensional exciton dynamics. This work provides direct experimental evidence that electronic delocalization can be decoupled from cooperative coherence in CdSe quantum dot superlattices.

When multiple emissive units interact coherently through a common light field, the system can exhibit collective optical behaviors that are different from those of individual emissive units[1, 2], namely cooperative emission. Owing to the emergence of macroscopic phase coherence and collective light-matter interactions among a large number of emitters[3, 4], cooperative emission provides a unique platform for exploring many-body quantum phenomena and developing quantum photonic technologies.

However, realizing these processes requires strict requirements for the material system. An ideal system usually needs a larger oscillator strength, small inhomogeneous broadening, and weak exciton dephasing[5]. Therefore, so far, experimental observations of cooperative emission have mainly been limited to a few highly controllable systems, such as CuCl nanocrystals[6] and KCl crystals[7]. In contrast, realizing cooperative emission in solution-processable materials remains a major challenge.

Quantum dot (QD) superlattices offer a highly promising approach to addressing this challenge. Their ordered periodic arrangement can reduce the impact of local structural disorder, while the delocalization of electron states may enhance the collective interactions between the emitting materials. Therefore, QD superlattices have become an important platform for exploring cooperative emission in solution-processing systems. The first observation of superfluorescence in $CsPbBr_3$ perovskite QD superlattices was reported by Rainò et al. in 2018 at 6 K[5]. Subsequent studies further expanded the achievable temperature range and structural diversity of cooperative emission in perovskite QD superlattices, reporting superfluorescence at 77 K[8, 9], 105 K[10], and even room temperature[11]. In 2022, Blach[12] et al. reported superradiance in perovskite QD superlattices through temperature-dependent PL linewidth and lifetime measurements. It is worth noting that in these studies, cooperative emission is usually accompanied by a steady-state PL feature with a distinct redshift spectrum. This spectral behavior is similar to that observed in CdTe and CdSe quantum dot superlattices in terms of quantum resonance emission[13-15]. Quantum resonance refers to the phenomenon where when the energy levels of electron transitions between quantum dots are close to each other and the distance between the dots is small enough to allow for significant overlap of wave functions, electrons may tunnel between adjacent quantum dots, forming a delocalized state across the quantum dots. Mechanistically, delocalization may enhance oscillator strength, reduce inhomogeneous broadening, etc., and quantum resonance may have a certain promoting effect on cooperative emission. Therefore, this similarity raises a fundamental question: does the appearance of a delocalized quantum resonance state necessarily imply the establishment of coherence and cooperative emission?

This question is nontrivial because delocalization and coherence, although related, are not equivalent. Delocalization describes the spatial extension of an excited state over multiple coupled sites, whereas coherence requires a well-defined phase relation to be established and maintained across those sites. To date, discussions on the distinction between delocalization and coherence have mainly focused on organic molecular aggregates and multimeric systems, and have largely been based on theoretical investigations[16-18]. Through molecular dynamics simulations, these studies have demonstrated that delocalization does not necessarily lead to coherence; rather, coherent emission is further constrained by factors such as static disorder, phonon scattering, and dark-state participation. However, direct experimental clarification of the relationship between delocalization and coherence remains limited. QD superlattices provide a promising platform for addressing this issue, where electronic delocalization can be engineered and correlated with collective optical responses.

Here, we build on our previous work, in which quantum resonance emission in CdSe QD superlattices was identified as a red-shifted Lorentzian-shaped emission arising from resonant interdot coupling and electronic delocalization. The disappearance of size-selective emission further confirmed that this emission is no longer governed by isolated-QD transitions, but originates from delocalized states extending across coupled QDs. Notably, CdSe QDs can exhibit a dephasing time of approximately 109 ps at 5 K[19], providing a relatively long coherence window in the cryogenic regime. These features make CdSe QD superlattices a promising platform for experimentally testing whether electronic delocalization is sufficient to establish cooperative radiative coherence. To address this question, we combine power-dependent and temperature-dependent spectroscopy measurements, which together probe the excitation-density dependence, radiative dynamics, and disorder- or phonon-related decoherence processes of the quantum resonance state. This approach allows us to examine whether a delocalized emission state develops the key signatures of cooperative radiation. The results experimentally distinguish delocalization from coherence in CdSe QD superlattices and show that delocalized quantum resonance emission does not necessarily evolve into cooperative radiative emission.

CdSe QD superlattices were fabricated by self-assembly at the gas-liquid interface[20-22]. Figure 1a shows the room-temperature PL spectra of the QD superlattices and QD solution, with the inset displaying a TEM image of the QD superlattice. The PL peak of the QD superlattice appears at around 630 nm. Compared with the QD solution, the QD superlattice emission exhibits a pronounced red shift (~133 meV) and a reduced full width at half-maximum (FWHM), narrowing from ~28 nm to ~20 nm, which are indicative of delocalized quantum resonance emission. For cooperative emission to emerge, a sufficiently high excitation density is required to establish a common radiation field that can drive phase-correlated electronic oscillations among coupled QDs. Power-dependent PL spectroscopy is therefore used to determine whether the delocalized quantum resonance state develops cooperative-emission signatures. Figure 1b shows the power-dependent PL spectra of the CdSe QD superlattices measured at cryogenic temperature. Figures 1c, 1d summarize the peak intensity, peak position, and FWHM of quantum resonance peak as a function of excitation power density.

As the excitation power density increases, the intensity of quantum resonance peak increases continuously and follows a sublinear power law of approximately $I \propto P^{0.77}$. Meanwhile, the emission peak gradually red-shifts and the spectral width broadens. The sublinear intensity scaling indicates that quenching or dissipative recombination remains active at low temperature. Although cryogenic conditions suppress most thermally activated nonradiative channels, the emission intensity does not show the superlinear enhancement expected for cooperative emission. The simultaneous red-shift and linewidth broadening suggest that local photothermal effects may present under pulsed excitation.

Cooperative emission is generally expected to exhibit power-dependent linewidth narrowing and superlinear growth of the emission peak intensity, arising from enhanced coherence and coherent superposition of the radiation fields from individual emitting dipoles. However, neither signature is observed in the present CdSe QD superlattices, indicating that the steady-state PL spectra provide no direct evidence for cooperative emission.

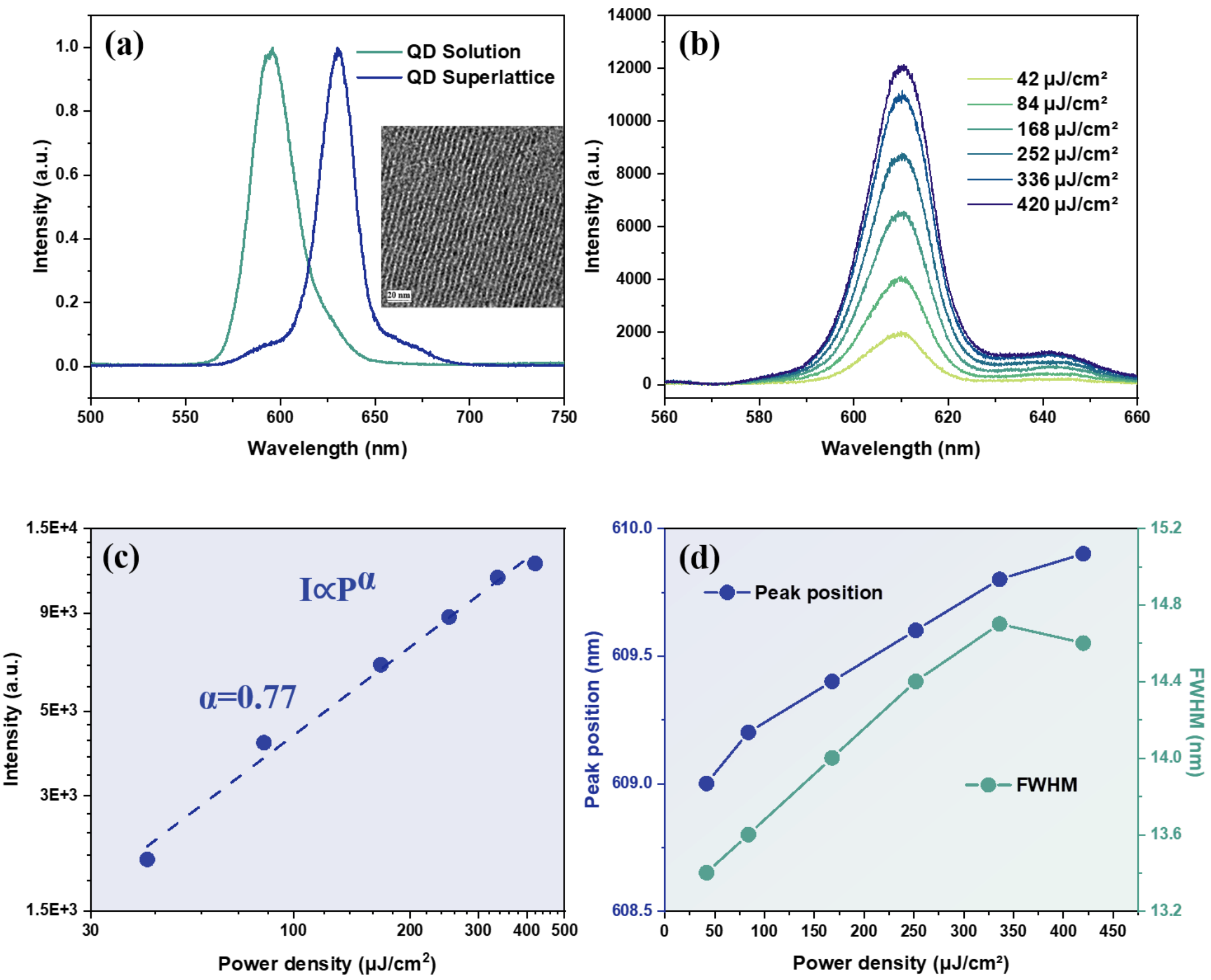

**Figure 1.** (a) Room-temperature PL spectra of the QD solution and QD superlattice. The inset shows a TEM image of the QD superlattice. (b) Power-dependent PL spectra of the QD superlattice measured at cryogenic temperature. (c) PL peak intensity as a function of excitation power density. (d) PL peak position and FWHM as a function of excitation power density.

To further examine whether the quantum resonance emission exhibits temporal signatures of cooperative emission, such as power-dependent emission delay or accelerated radiative decay, power-dependent time-resolved PL measurements were performed at cryogenic temperature. The decay curves of the quantum resonance emission are shown in Figure 2a. At low excitation density, the integrated emission intensity increases nearly linearly with power. At higher excitation density, the power-law exponent decreases to approximately $I \propto P^{0.56}$ (Figure 2b), indicating that nonradiative recombination becomes increasingly important under high-density excitation.

The decay curves of the quantum resonance peak can be fitted using a biexponential model. The evolution of the fitting parameters is summarized in Figure 2c. With increasing excitation power, A1 increases, A2 decreases, τ1 shortens, and τ2 becomes longer. Meanwhile, the amplitude-weighted average PL lifetime decreases overall with increasing power density (Figure 2d). The fast component τ1 (~1.2 ns) can be attributed to multiexciton- or charged-state-related dissipative recombination[23]; its shortening, together with the increasing contribution of A1, is consistent with the activation of a faster loss channel at higher excitation density. The slower component τ2 (~5.6-8.8 ns) is more likely associated with the neutral exciton radiative channel. As the excitation density increases at low temperature, trap-state filling could partially suppress nonradiative trapping pathways, resulting in an apparent lengthening of the lifetime associated with the neutral exciton channel. At the same time, the increased contribution of a fast dissipative channel likely dominates the ensemble-averaged decay dynamics, leading to a decrease in the weighted average lifetime[24].

Low-temperature time-resolved PL measurements of the Coulomb-coupled peak are shown in Figure S1. Compared with this peak, the quantum resonance peak exhibits a higher power-law exponent and a longer average lifetime throughout the measured power range. This comparison indicates that, under same temperature and excitation conditions, the quantum resonance emission channel is less strongly affected by defect trapping and nonradiative dissipation. This enhanced defect tolerance can be understood as a consequence of wave-function delocalization induced by strong electronic coupling, which reduces the probability that carriers are captured by a single localized defect and weakens the influence of surface states on the recombination dynamics.

However, the quantum resonance peak still does not display the characteristics of cooperative emission. No pronounced shortening of the average PL lifetime, delayed emission, or superlinear growth of the emission intensity is observed. Thus, although low temperature can extend the coherence retention time and create more favorable conditions for phase correlation, the present quantum resonance emission does not show the dynamical behavior expected for cooperative radiation.

These observations demonstrate that electronic delocalization provides a physical basis for collective optical behavior, but does not by itself establish macroscopic coherence. Delocalization describes the spatial extension of an excited state determined by interdot electronic coupling, whereas coherence depends on whether phase correlations among different sites can survive static disorder, residual phonon coupling, and environmental fluctuations. This interpretation is consistent with theoretical studies of organic molecular aggregates and multimeric systems[16-18], which have shown that delocalization can exist independently of cooperative emission, whereas cooperative radiation requires more stringent dynamical stability.

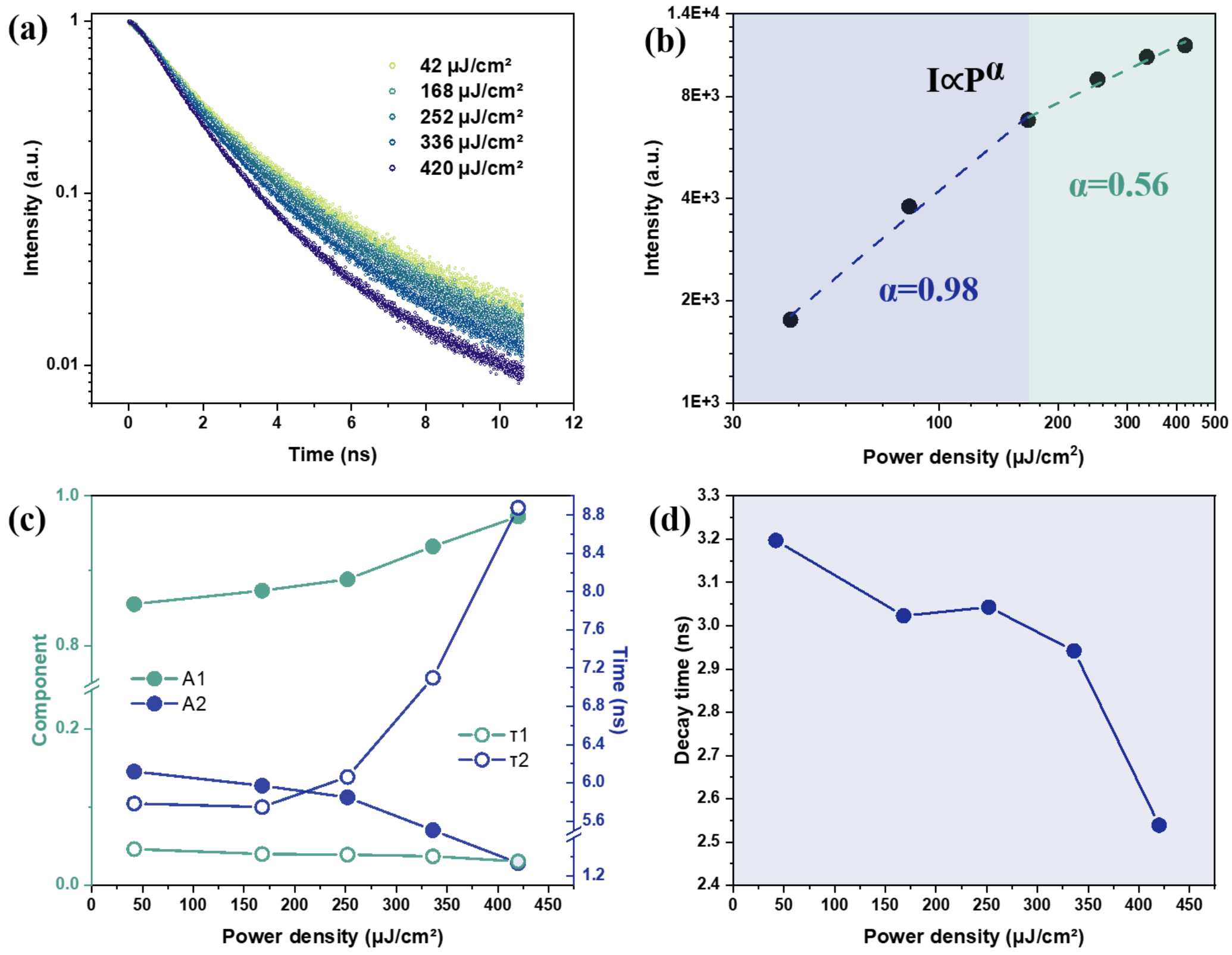

**Figure 2.** (a) Power-dependent time-resolved emission intensity decay curves of the quantum resonance peak. (b) Evolution of intensity with power density. (c) Evolution of the fitted lifetime components and amplitudes with power density. (d) Evolution of averaged decay time with power density.

To explore the possible factors that limit the formation of coherent cooperative emission, temperature-dependent steady-state PL spectroscopy was performed (Figure 3). The temperature evolution of the peak position, linewidth, and emission intensity provides insight into disorder- and phonon-related dephasing processes, as well as dark-exciton participation, all of which can hinder the establishment of macroscopic phase coherence. Figure 3a shows the temperature-dependent PL spectra, and Figures 3b-d summarize the extracted peak position, FWHM, and peak intensity. As shown in Figure 3b, the quantum resonance peak shifts to lower energy with increasing temperature. For a direct-bandgap semiconductor, the temperature dependence of the exciton energy is commonly described by the Varshni equation:

$$E(T) = E_0 - \frac{\alpha T^2}{T + \beta} \tag{1}$$

where $E_0$ is the low-temperature exciton energy, $\alpha$ is the temperature coefficient, and $\beta$ is an empirical parameter related to lattice vibrations. The dashed curve in Figure 3b was calculated using bulk CdSe parameters[25] ($\alpha = 3.7 \times 10^{-4}$ eV/K，$\beta$=150 K). The experimentally observed peak-position evolution deviates markedly from the bulk CdSe prediction, with the deviation becoming more pronounced at lower temperatures. In contrast to weakly coupled QD films, where the deviation from the bulk trend is typically close to a constant offset[26], the distinct temperature dependence observed here further supports the formation of a strongly coupled quantum resonance state[27].

Figure 3c shows the temperature dependence of the FWHM of the quantum resonance peak. The linewidth increases monotonically as the temperature rises. For semiconductor nanostructures, the temperature-dependent linewidth is commonly expressed as:

$$\Gamma(T) = \Gamma_{inh} + \sigma * T + \Gamma_{LO} * \left(\exp\left(\frac{E_{LO}}{k_B T}\right) - 1\right)^{-1} \tag{2}$$

where $\Gamma_{inh}$ is the temperature-independent inhomogeneous broadening, which mainly originates from variations in QD size, morphology, composition, and local dielectric environment; $\sigma$ is the exciton-acoustic phonon coupling coefficient; $\Gamma_{LO}$ is the exciton-LO phonon coupling coefficient; $k_B$ is the Boltzmann constant; and $E_{LO}$ is the longitudinal optical phonon energy of CdSe QDs, taken as 25.6 meV. The fitted parameters obtained from Equation (2) are compared with values reported for other CdSe systems in Table 1.

**Table 1.** Linewidth parameters obtained from temperature-dependent FWHM fitting for CdSe-based systems.

| Sample | $\Gamma_{inh}$ (meV) | $\sigma$ (μeV/K) | $\Gamma_{LO}$ (meV) |
|---|---|---|---|
| **CdSe QD superlattice (This work)** | 50.5 | 39.7 | 8.8 |
| **CdSe QDs[28]** | 133.3 | 85.4 | 27.4 |
| **Bulk CdSe[29]** | — | 0.06 | 18.3 |
| **CdSe thin film[30]** | 5.5 | 84 | 20 |
| **CdSe nanoplatelets[31]** | 32.5 | 9.8 | 11.6 |

As shown in Table 1, $\Gamma_{inh}$ for the present QD superlattices is much smaller than that of disordered CdSe QD ensembles, i indicating that self-assembly suppresses statistical variations in QD size distribution and local dielectric environment, thereby improving the overall uniformity of the system. Nevertheless, $\Gamma_{inh}$ remains the dominant contribution to the linewidth, suggesting that static disorder is still significant. The acoustic-phonon coupling coefficient $\sigma$ is much larger than that of bulk CdSe, which can be attributed to enhanced sensitivity of confined excitons to lattice deformation[32]. However, $\sigma$ is lower than values typically reported for colloidal QDs and disordered thin films, suggesting that the present system cannot be simply regarded as an ensemble of isolated zero-dimensional emitters. Instead, interdot coupling and exciton delocalization may partially suppress acoustic-phonon scattering. The relatively small $\Gamma_{LO}$ value, governed by the Frohlich interaction, may also originate from delocalization-induced reduction of the effective polarization strength[33].

Although weaker phonon scattering is favorable for coherence retention, the large $\Gamma_{inh}$ indicates that static disorder continues to disrupt energy matching and phase synchronization among emitting units[34]. Therefore, linewidth analysis suggests that static-disorder-dominated dephasing is a key factor limiting the establishment of cooperative emission. Low-temperature continuous-wave PL spectrum further reveals distinct spectral features (Figure S2): multiple sub-peaks appear near the quantum resonance main peak and become less distinguishable as the probed area increases. These sub-peaks can be attributed to superlattice domains with different orientations or coupling strengths[35]. Variations in ordering, the number of QDs participating in coupling, local coupling strength, and energy-level structure can lead to resolvable shifts in the emission peak position. The main peak therefore represents an ensemble-averaged emission from dominant domain structures within the probed region. The fitted linewidths of individual sub-peaks are only on the order of ~10 meV, much narrower than the ensemble linewidth, indicating that inter-domain spectral heterogeneity contributes substantially to the observed static disorder.

The peak intensity of the quantum resonance emission exhibits a nonmonotonic temperature dependence (Figure 3d). As temperature increases, the PL intensity first increases and then decreases. This anomalous behavior can be understood in terms of population redistribution between dark and bright exciton states[36]. At low temperature, band-edge excitons predominantly occupy weakly radiative dark states. With increasing temperature, dark excitons can be thermally activated into bright exciton states, thereby enhancing radiative recombination and increasing the PL intensity. At higher temperatures, nonradiative recombination channels and phonon scattering become more pronounced, leading to thermal quenching and a subsequent decrease in emission intensity[37, 38].

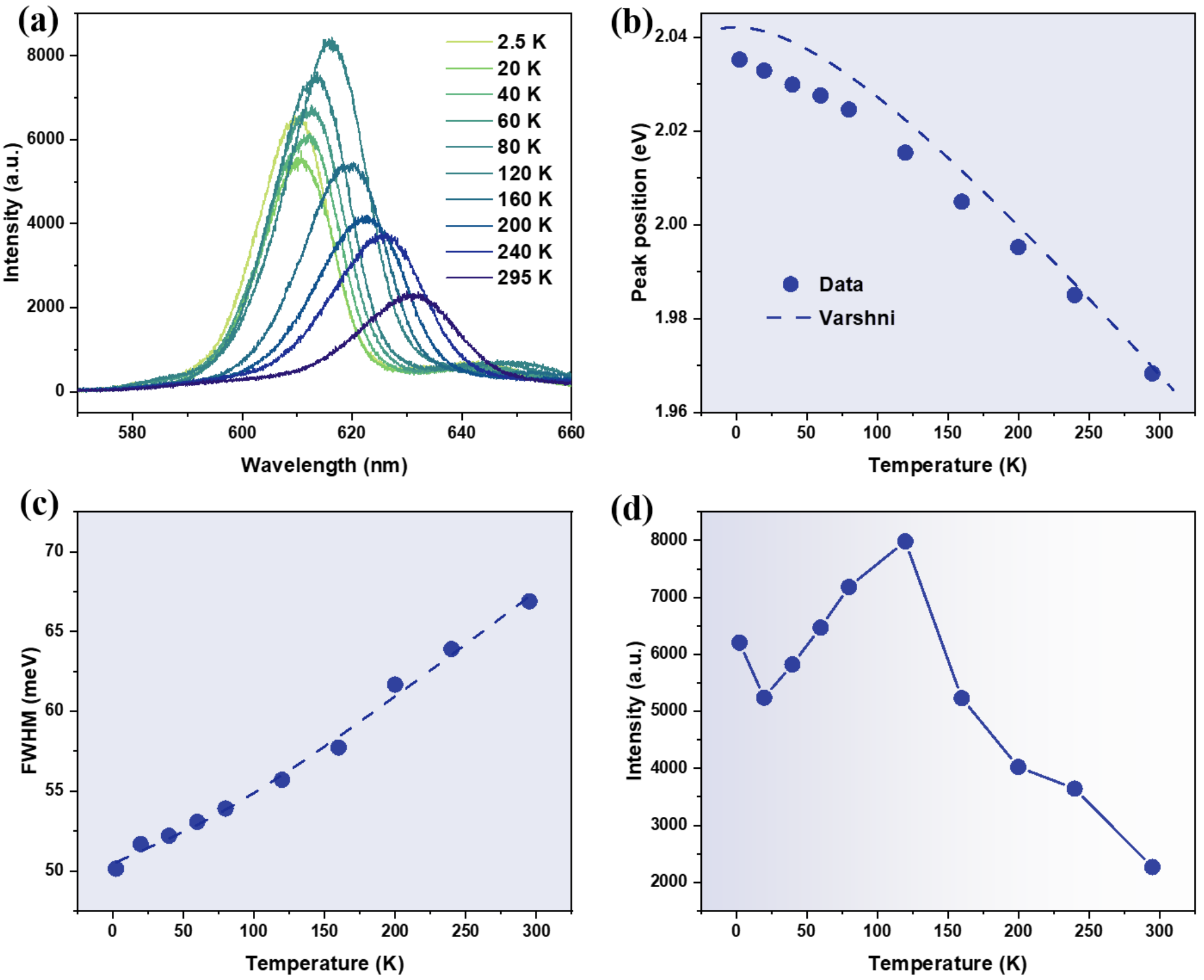

**Figure 3.** (a) Temperature-dependent PL spectra of the QD superlattices. (b) Temperature dependence of the peak position. (c) Temperature dependence of the FWHM. (d) Temperature dependence of the peak intensity.

Overall, the temperature-dependent steady-state PL results point to two main factors that limit cooperative emission in CdSe QD superlattices. First, the system contains pronounced multiscale static disorder, including intra-domain local structural fluctuations and inter-domain heterogeneity. Second, dark-exciton occupation at cryogenic temperatures reduces the effective oscillator strength and weakens macroscopic dipole polarization. Together, these mechanisms suppress the establishment of phase synchronization among emitting units and explain why cooperative emission remains absent even under low-temperature conditions.

To further clarify the temperature-dependent dynamics of quantum resonance emission, time-resolved PL measurements were performed at different temperatures (Figure 4). The decay curves of the quantum resonance peak were fitted using a biexponential function, yielding a fast component $\tau 1$ (~1.2 ns), a slower component $\tau 2$ (~7.5-10 ns), and their amplitude weights, A1 and A2. As the temperature increases, A1 gradually decreases and A2 increases, accompanied by a slight shortening of both $\tau 1$ and $\tau 2$. Despite the shortening of the individual lifetime components, the amplitude-weighted average lifetime increases overall with temperature.

This behavior is consistent with a temperature-induced redistribution of exciton populations among different recombination channels. The slight shortening of $\tau 1$ and $\tau 2$ can be attributed to enhanced exciton–phonon interactions and thermally activated nonradiative processes at elevated temperatures[39, 40]. Meanwhile, thermal activation may promote carrier escape from localized trap states or weakly emissive states and redistribute carriers into the main radiative exciton channel. However, thermal activation can also promote carrier escape from localized trap states and redistribute carriers into the radiative exciton channel. As a result, the contribution of the fast dissipative channel decreases[41], whereas the weight of the slower emissive channel increases. Because τ2 is much longer than τ1, the increase in A2 dominates the ensemble-averaged decay dynamics, leading to an overall lengthening of the amplitude-weighted average lifetime within the measured temperature range.

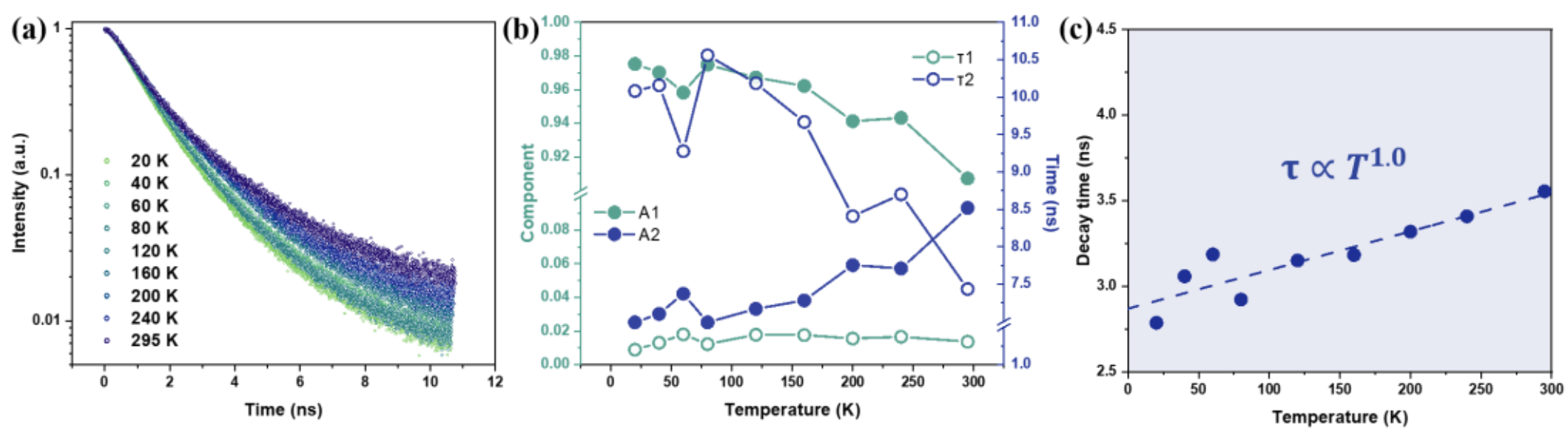

**Figure 4.** (a) Temperature-dependent time-resolved emission intensity decay curves of the quantum resonance peak. (b) Evolution of the fitted lifetime components and amplitudes with temperature. (d) Evolution of averaged decay time with temperature.

The temperature dependence of the principal radiative channel was further analyzed using the radiative-lifetime model for low-dimensional excitons. In an ideal crystal, momentum conservation allows only excitons within the radiative window near the Brillouin-zone center ($K \approx 0$) to recombine directly. In real systems, however, phonon scattering gives the free-exciton emission a finite linewidth $\Delta$, so excitons over a finite momentum range can contribute to radiative emission[42]. As temperature increases, thermal equilibrium drives a larger fraction of excitons into higher-momentum states, reducing the fraction that remains within the radiative window[43]. This process can be expressed as:

$$\tau(T) = \tau_0 \frac{1}{\zeta(T)} \tag{3}$$

$$\zeta(T) = \frac{\int_0^{\Delta} D_{ex}(E) e^{-\frac{E}{k_B T}} dE}{\int_0^{\infty} D_{ex}(E) e^{-\frac{E}{k_B T}} dE} \tag{4}$$

where $\tau_0$ is the PL decay time at 0 K, $\zeta(T)$ is the fraction of excitons that contributes to PL emission, $D_{ex}(E)$ is the exciton density of states, and $k_B$ is the Boltzmann constant. Substituting the analytical density of states for one-dimensional and two-dimensional excitons gives different power-law relationships for the average radiative lifetime:

$$\tau(T) = \begin{cases} \tau_0 \sqrt{\frac{\pi k_B T}{4\Delta}} \ (1\text{D}) \\ \tau_0 \frac{k_B T}{\Delta} \ (2\text{D}) \end{cases} \tag{5}$$

Fitting the data in Figure 4c shows that the PL decay time of the quantum resonance peak follows an approximately $\tau(T) \propto T^{1.0}$ dependence. This behavior is consistent with two-dimensional exciton dynamics, indicating that the quantum resonance peak does not originate from isolated zero-dimensional emitters. Instead, it arises from an excitonic state spatially extended within locally ordered regions of the superlattice, with significant in-plane delocalization.

In conclusion, using CdSe QD superlattices as a model system with experimentally established quantum resonance delocalization, this work demonstrates that electronic delocalization alone is not sufficient to produce coherent cooperative emission. Power-dependent steady-state and time-resolved PL measurements reveal that the quantum resonance peak has a longer lifetime and greater defect tolerance than the weakly coupled emission channel. The temperature dependence of the decay lifetime follows two-dimensional exciton dynamics, indicating in-plane delocalization within ordered superlattice domains. However, the absence of superlinear intensity scaling, emission delay, and radiative-lifetime acceleration rules out superfluorescence- or superradiance-like behavior under the present conditions. Temperature-dependent spectroscopy further identifies the key factors that suppress macroscopic coherence: substantial static disorder, domain-to-domain spectral heterogeneity, and low-temperature dark-exciton population. These effects disrupt energy matching and phase synchronization among coupled emitting units, even though the emissive state remains spatially extended. Overall, these results experimentally decouple quantum-resonance delocalization from cooperative coherence and suggest that future cooperative emission from colloidal QD superlattices will require stronger structural uniformity, reduced static disorder, and suppressed dark-state losses.

ASSOCIATED CONTENT

**Supporting Information**.

Details of sample fabrication, experimental methods, PL spectrum of CdSe QD superlattice, Power-dependent time-resolved emission intensity decay curves of the Coloumb coupling peak, PL spectrum of CdSe QD superlattice under continuous-wave laser and TEM images of superlattices.

AUTHOR INFORMATION

**Corresponding Author**

Butian Zhang-National Gravitation Laboratory, MOE Key Laboratory of Fundamental Physical Quantities Measurement, and School of Physics, Huazhong University of Science and Technology, Wuhan 430074, People's Republic of China; orcid.org/0000-0002-9959-3912; Email: butianzhang@hust.edu.cn

**Authors**

Lanfang Hou-National Gravitation Laboratory, MOE Key Laboratory of Fundamental Physical Quantities Measurement, and School of Physics, Huazhong University of Science and Technology, Wuhan 430074, People's Republic of China

Zijian He-Wuhan National Laboratory for Optoelectronics and School of Physics, Huazhong University of Science and Technology, Wuhan 430074, China

Kexin Wang- Yunnan Key Laboratory of III-V Infrared Detectors, Kunming Institute of Physics, Kunming 650223, China

Kai Wang-Wuhan National Laboratory for Optoelectronics and School of Physics, Huazhong University of Science and Technology, Wuhan 430074, China

Shun Wang-National Gravitation Laboratory, MOE Key Laboratory of Fundamental Physical Quantities Measurement, and School of Physics, Huazhong University of Science and Technology, Wuhan 430074, People's Republic of China; orcid.org/ 0000-0001-6890-8371; Email: shun@hust.edu.cn

**Author Contributions**

The manuscript was written through contributions of all authors. All authors have given approval to the final version of the manuscript.

**Notes**

The authors declare no competing financial interest.

*Supporting Information*

# Electron Delocalization versus Emission Coherence of Quantum Dot Superlattices

*Lanfang Hou[1], Zijian He[2], Kexin Wang[3], Kai Wang[2], Shun Wang[1], Butian Zhang[1,*]*

[1]National Gravitation Laboratory, MOE Key Laboratory of Fundamental Physical Quantities Measurement, and School of Physics, Huazhong University of Science and Technology, Wuhan 430074, People's Republic of China

[2]Wuhan National Laboratory for Optoelectronics and School of Physics, Huazhong University of Science and Technology, Wuhan 430074, China

[3]Yunnan Key Laboratory of III-V Infrared Detectors, Kunming Institute of Physics, Kunming 650223, China

**Corresponding author(s). E-mail(s): butianzhang@hust.edu.cn*

**METHODS**

**Growth of CdSe QD superlattices.** The CdSe QDs are purchased from Mesolight Inc (Suzhou, China). The superlattice is prepared at the gas-liquid interface in oxygen-free glove box. The solution of CdSe colloidal QDs is diluted to 25 mM in n-octane or toluene solutions. Subsequently, 20 μl of the QD solution is dropped onto the upper surface of 10 ml of N,N-Dimethylformamide (DMF) in a polytetrafluoroethylene container. The top of the container is covered with an aluminum foil to slow the evaporation of the octane solvent. The liquid phase in the container was left undisturbed for at least 6 hours.

**Optical spectroscopy.** Steady-state PL spectra are obtained by using a 488 nm laser, a Princeton Instruments SP-2500 spectrometer, and a Pixis 100 CCD camera. The time-resolved PL spectra are collected by using the FLINT8 APE laser with an excitation wavelength of 515 nm and a pulse width of 120 fs at a repetition frequency of 76 MHz.

**TEM Characterization.** Transmission electron microscopy images are performed on an FEI Tecnai G2 F30. Operating at 200 kV.

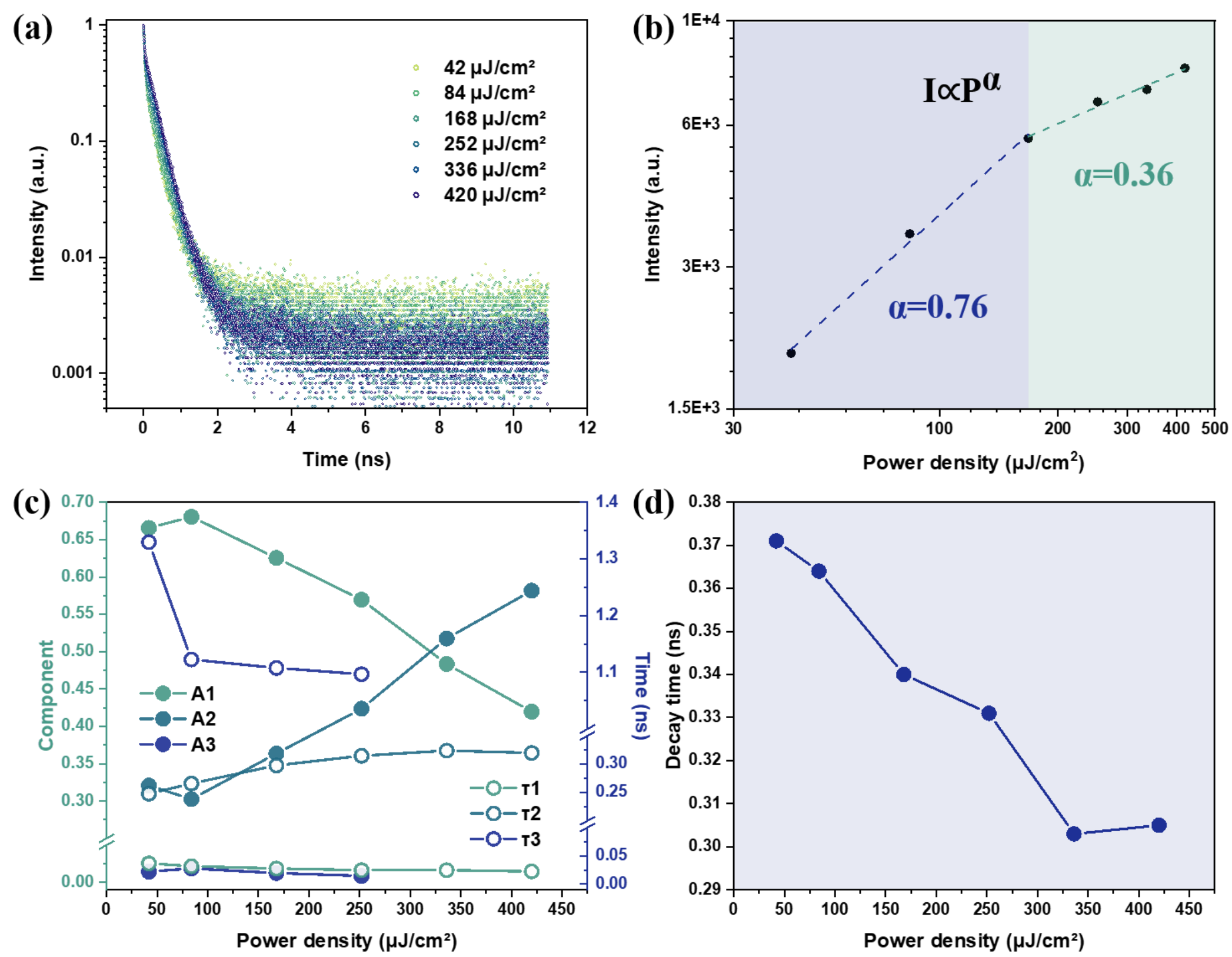

**Figure S1.** (a)Power-dependent time-resolved emission intensity decay curves of the Coloumb coupling peak. (b) Evolution of intensity with power density. (c) volution of the fitted lifetime components and amplitudes with power density. (d) Evolution of averaged decay time with power density.

After fitting the decay profiles of the Coulomb-coupled emission peak with a multiexponential function, three lifetime components, τ1, τ2, and τ3, together with their corresponding amplitude weights, A1, A2, and A3, were extracted. With increasing excitation power, A1 and A3 gradually decrease, whereas A2 increases. Meanwhile, τ1 remains nearly unchanged, τ2 increases monotonically, and τ3 decreases monotonically. As a result, the weighted average lifetime shows an overall decrease with increasing excitation power.

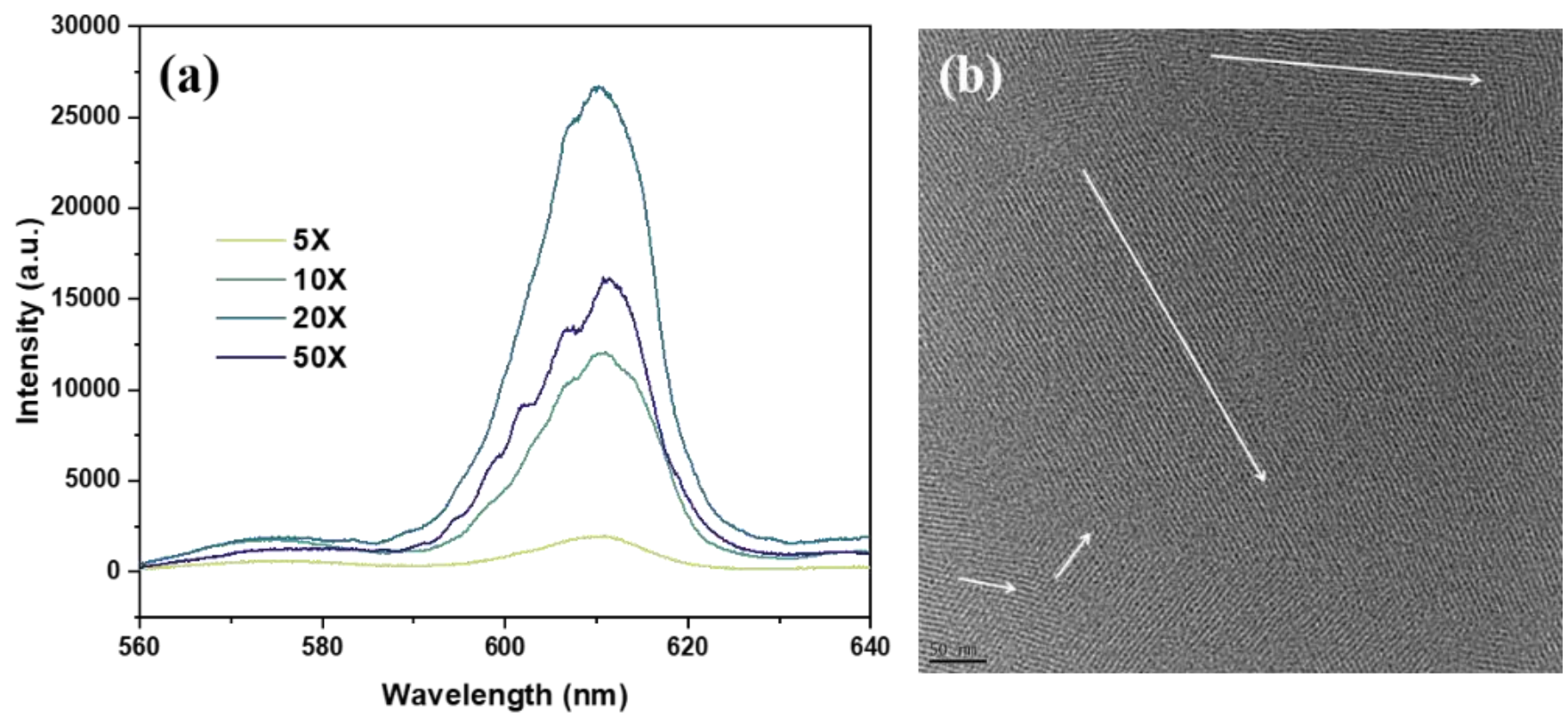

**Figure S2.** (a)The PL spectra of the sample under continuous-wave laser excitation at 532 nm under different magnification lenses at 2.8 K. (b) The TEM image of the sample at a large field of view.

Under the 2.8 K continuous-wave excitation condition, a series of sub-peaks appeared near the main peak of quantum resonance, and these sub-peaks gradually disappeared as the detection range expanded (the objective lens magnification decreased). The TEM image observation results showed that multiple domains with different orientations appeared within the approximate micrometer-sized range.